%\magnification= 1200
\hsize 12.5truecm
\vsize 19truecm
%  The following 13 lines establish the use of the Euler Fraktur font.
%
\font\teneuf=eufm10 at 12pt
\font\seveneuf=eufm7 at 9pt
\font\fiveeuf=eufm5 at 7pt
\newfam\euffam
\textfont\euffam=\teneuf
\scriptfont\euffam=\seveneuf
\scriptscriptfont\euffam=\fiveeuf
\def\frak{\relaxnext@\ifmmode\let\next\frak@\else
\def\next{\Err@{Use \string\frak\space only in math mode}}\fi\next}
\def\goth{\relaxnext@\ifmmode\let\next\frak\else
\def\next{\Err@{Use \string\goth\space only in math mode}}\fi\next}
\def\frak#1{{\fam\euffam#1}}
\def\frakk#1{\noaccents@\fam\euffam#1}
%  End definition of Euler Fraktur font.

\font\tit=cmbx12 scaled\magstep 1
\font\abst=cmsl9

\def\centra#1{\vbox{\rightskip=0pt plus1fill\leftskip=0pt plus1fill #1}}

\long\def\summary#1{\bigskip\medskip
\vbox{\par\leftskip=30truept\rightskip=30truept
\noindent{\bf Summary:} \abst #1}\parindent=15truept}

%             Diagrammi

\def\mapright#1{\smash{\mathop{\longrightarrow}\limits^{#1}}}

\def\mapdown#1{\Big\downarrow\rlap
 {$\vcenter{\hbox{$\scriptstyle#1$}}$}}

\hbox{ }
\vskip 3truecm
\centra{ \tit A geometric definition of Lie derivative\goodbreak for Spinor
Fields}
\vskip 10pt
\centra{{ }Lorenzo FATIBENE,
  Marco FERRARIS,\goodbreak Mauro FRANCAVIGLIA and
  Marco GODINA}
\vskip 10pt
\centra{Istituto di Fisica Matematica ``J.--L. Lagrange''\goodbreak
Universit\`a di Torino, Via C. Alberto 10, 10123 TORINO (ITALY)}

%%%%%%%%%%%%%\maketitle

%            Definizioni Matematiche

\def\Re{I \kern-.36em R}              %  Numeri reali
\def\Co{I \kern-.66em C}              %  Numeri complessi
                     %  Derivata parziale
                       %  Nabla
\def\rdot{{\raise.5ex \hbox{.}}}      %  Moltiplicazione
                 %  Infinito
                   %  Intersezione
                  %  Unione disgiunta
                   %  Unione
\def\ida{\hbox{\kern .07em}}		  %  Indice diagonale A^{i}\ida_{j}

\def\det{\hbox{\rm det}}
\def\Cl{\hbox{\it Cl}}
\def\Spin{\hbox{\it Spin}}
\def\Pin{\hbox{\it Pin}}
\def\Aut{\hbox{\it Aut}}

                                      %  oppure A_{i}\ida^{j}
%

\def\spin{{\frak s\frak p\frak i\frak n}(p,q)}		% algebra di spin
\def\so{{\frak s\frak o}(p,q)}		  														% algebra del gruppo SO(p,q)

\def\Real{I \!\! R}

\def\ba{\begin{array}}
\def\ea{\end{array}}

\footnote{}{This paper is in final form and no version of it will be
submitted for publication
 elsewhere.}

\summary{
Relying on the general theory
of Lie derivatives a new geometric
definition of Lie derivative for general spinor fields is given,
more general than Kosmann's one. It is shown that for particular
infinitesimal lifts, i.e. for Kosmann vector fields, our definition coincides
with the definition given by Kosmann more than 20 years ago.
}

%keywords{$G$--bundle, connection, spin structure}
%Classification. 53C25; 19L10, 58G10.
%%%%\end{abstract}

\vskip5pt\noindent
{\bf 1 Introduction}

It is known that starting from some works by physicists (Dirac [1,2], Pauli [3])
the notion of spinor was introduced in Physics, although
its discovery is due to E. Cartan in 1913 (cf. [4]) in his researches on the
linear  representations of some simple Lie groups.

\noindent A pretty nice and detailed algebraic theory was given later by
C. Chevalley [5] and other papers on Clifford--algebras and spinor groups were
published, in particular by J. Dieudonn\'e [6], Atiyah, Bott, Shapiro [7],
Milnor [8].

Spinors on curved space--time manifolds were introduced and used
since 1928 by physicists.
In a series of very important papers Yvette Kosmann [9,10,11,12] introduced the
notion of Lie derivative for spinor fields on spin manifolds and the study
of the
problem of transforming a spinor field under one--parameter group of
diffeomorphisms of the base manifold.
\vskip 5 pt
\noindent However, thanks to her beautiful and relevant work, we have the
feeling that Kosmann's theory was not formulated in the clearest possible way
and that the geometric meaning of her {\it ``ad hoc"}
definition was still lacking.
In fact, it is quite known that the difficulty in defining the Lie derivative
of a spinor field on a space (and time) oriented (pseudo) Riemannian
manifold $(M,g)$ comes from the fact
that there is no natural definition of the image of such a spinor field by
a diffeomorphism. This corresponds to the fact that the bundle $SO(M,g)$
together with any spin bundle $\Spin(M,g)$ covering $SO(M,g)$
are not natural bundles [17].
\vskip5pt
\noindent Of course, these bundles are particular gauge--natural
bundles [17,21] and we will show
moreover that it is always possible to lift in a unique way any vetor field
$\xi \in {\frak X}(M)$ of the base manifold $(M,g)$.

The corresponding vector field
$\tilde \xi_{K} \in {\frak X}(\Spin(M,g))$, projectable over $\xi$,
and $\xi_{K} \in{\frak X}(SO(M,g))$ also projectable over $\xi$,
will be called the Kosmann vector fields of
$\xi \in {\frak X}(M)$. We remark that the vector field $\tilde \xi_{K}$
also projects over $\xi_{K}$.

The Kosmann lifting is not natural and differs from the Levi--Civita
lifting, i.e. the principal lifting induced by the Levi--Civita connection
on $SO(M,g)$ (first order derivatives of $\xi$ are involved),
although the Ricci rotation coefficients of the Levi--Civita connection
appear in the expression which gives the
Kosmann lifting of $\xi$.
\vskip5pt
In this paper we shall give a {\it new}
definition of Lie derivative for spinor fields
and we will show that for particular
infinitesimal lifts, i.e. for Kosmann vector fields, our definition coincides
with the definition given by Kosmann more than 20 years ago.

\vskip5pt\noindent
{\bf 2 Spin structures}

Let $(P,M,\pi,G)$ be a principal fiber bundle over $M$ with
structural group a Lie group $G$. Let $\rho \colon \Gamma \rightarrow G$ be
a central homomorphism of a Lie group  $\Gamma$ onto $G$ with kernel $K$, that
is $K$ is discrete and contained in the center of $\Gamma$ (see Greub and
Petry in [13]; see also [22]).
We recall that a $\Gamma$--{\it structure} on $P$ is a principal bundle map
$\eta \colon \tilde P \rightarrow P$ which is equivariant under the right
actions of the structure groups, that is:

$$
\eta (\tilde z \cdot \gamma) = \eta (\tilde z) \cdot \rho (\gamma) \quad ,
\tilde z \in \tilde P, \; \gamma \in \Gamma \, ,
\eqno (2.1)
$$

\noindent and where $\rho$ is assumed central by definition.

\noindent Equivalently the following diagram commutes:

%%%%%\begin {array}{ccc}
\vskip 5pt
$$
\matrix{
\tilde P \times \Gamma & \mapright{\eta \times \rho}&P \times G \cr
\mapdown{\tilde R } &  & \mapdown{ R} \cr
\tilde P & \mapright{\eta} & P \cr
\mapdown{\tilde \pi}  &  & \mapdown{ \pi} \cr
M & \mapright{id_{M}} & M \cr
}
\eqno (2.2)
$$
\vskip 5pt

\noindent where $R$ and $\tilde R$ denote right multipication in $P$
and $\tilde P$ respectively .

\noindent This means that for $\tilde z \in \tilde P$, both $\tilde z$ and
$\eta(\tilde z)$ lie over the same point, and that $\eta$ restricted to
a fiber is a ``copy" of $\rho$, i.e. equivalent to $\rho$.

In general, it is not guaranteed that a manifold admits a
spin structure. The existence condition for a $\Gamma$--structure on $P$ can be
formulated [13,14,22] in terms of {\v C}ech cohomology.

Let us also recall that for any principal fiber bundle
$(P,M,\pi,G)$ an {\it (principal) automorphism} of $P$ is a diffeomorphism
$\phi \colon P \rightarrow P$ such that
$\phi (u \cdot g) = \phi (u) \cdot g$,
for every $u \in P, \; g \in G$.
Each $\phi$ induces a unique
diffeomorphism $\varphi \colon M \rightarrow M$
such that $\pi \circ \phi = \varphi \circ \pi$. Accordingly, we denote by
$\Aut(P)$ the group of all principal (i.e. equivariant) automorphisms of $P$.
Assume a vector field $X \in {\frak X}(P)$ on $P$ generates
the one--parameter group $\phi_{t}$. Then, $X$ is $G$--invariant
if and only if $\phi_{t}$ is an automorphism of $P$ for every $t \in \Real$.
Accordingly, we denote by ${\frak X}_{G}(P)$
the Lie algebra of $G$--invariant vector fields of $P$.

Let $\Real^{p,q}$ be the familiar vector space $\Real^{n}$  ($p + q = n$)
equipped with the canonical nondegenerate symmetric bilinear form $<,>$
of signature $(p,q)$, i.e. with $p \,$ ``+'' signs and with $q \,$ ``-''
signs. We will
denote by $O(p,q)$ the (pseudo)--orthogonal group with respect to $<,>$,
that is $O(p,q) = \{L \in Gl(\Real^{p,q}) |  <Lu,Lv> = <u,v>\}$,
by $SO(p,q) = \{L \in O(p,q) | \, \det(L) = 1\}$ its special subgroup,
and by $SO_{0}(p,q)$ its connected component with identity. In the Euclidean
case ($p$ or $q = 0$), $SO(p,q) = SO_{0}(p,q)$ but in the general case they
are not equal.
To define the Clifford algebra $\Cl(p,q) \equiv \Cl(\Real^{p,q})$ we choose
any orthonormal basis $e_{a}$ of $\Real^{p,q}\subset \Cl(p,q)$,
that is $<e_{a},e_{b}> = \eta_{ab}$.
The Clifford algebra $\Cl(p,q)$ is the real vector space endowed
with an associative product generated (as an algebra)
by the unit $I$ and the elements
$e_{a}$, $1 \leq a \leq n=p+q$, with the relations:

$$
e_{a}e_{b} + e_{b}e_{a} =
-2\eta_{ab}I
\eqno (2.3)
$$

\noindent where $\eta_{ab} = 0$ if $a \not= b$, $+1$ if
$a = b \leq p$, and $-1$ if $p < a = b$. As a real vector
space, the Clifford algebra $\Cl(p,q)$ has dimension $2^{n}$.

\noindent The set $\{I, e_{a}, e_{a_{1}}e_{a_{2}},
..., e_{a_{1}}e_{a_{2}}\cdots e_{a_{r}},
e_{a_{1}} \cdots e_{a_{n}}:
a_{1} < a_{2} < \cdots < a_{r}\}$ forms a basis  of $\Cl(p,q)$.
The complexified Clifford algebra $\Co l(p,q)$ is the vector space $\Cl(p,q)
\otimes _{\Real}{\Co}$,  that is the vector space over the complex numbers
generated by the previous  elements $I, e_{a}, ...,
e_{a_{1}} \cdots e_{a_{n}}$.
In this paper we shall assume
that the dimension of $\Real^{n}$ is even, that is $n = 2d$. In this case it can
be proved that $\Co l(p,q)$ is isomorphic to the complex algebra
$M_{2^{d}}({\Co})$
of complex matrices of order $2^{d}$. The matrices
representing the fundamental elements $e_{a}$ will be denoted
by $\gamma_{a}$. They automatically carry a representation of various
subgroups of $\Cl(p,q)$ and in particular of $\Spin(p,q)$.

Let us briefly recall how $\Spin(p,q)$ is defined.
Consider the multiplicative group of units of $\Cl(p,q)$,
which is defined to be the subset

$$
\Cl^{\times}(p,q) \equiv  \{ a \in \Cl(p,q) \, : \, \exists a^{-1} \,
\hbox{ such that} \, a^{-1}a = aa^{-1} = 1 \} \quad .
\eqno (2.4)
$$

\noindent This group contains all elements
$v \in \Real^{p,q}\subset \Cl(p,q)$ with $<v,v>\not= 0$
so it is possible to consider the subgroup generated by its
``units length" elements,
$v \in \Real^{p,q}\subset \Cl(p,q)$ with $<v,v>= \pm 1$.

This subgroup it is denoted by $\Pin(p,q)$.
The spin group $\Spin(p,q)$ is defined as the subgroup of $\Pin(p,q)$
consisting of even elements:

$$
\Spin(p,q) = \Pin(p,q) \cap \Cl^{0}(p,q) \quad ,
\eqno (2.5)
$$

\noindent where $\Cl^{0}(p,q)$ is a subalgebra of $\Cl(p,q)$
called the {\it even part} of $\Cl(p,q)$.

\vskip5pt
Now, consider a space (and time) oriented (pseudo)--Riemannian manifold
with metric $g$ and dimension $n=dim(M)$.
In this case we have $G = SO(p,q)$,
$\Gamma = \Spin(p,q)$, $P = SO(M,g) \subset L(M)$, $\tilde P = \Spin(M,g)$
where $\Spin(p,q)$ is the usual notation for $\Spin(\Real^{p,q})$ and
$(L(M), Gl(n),M)$ is the principal bundle of linear frames.
We shall say that a {\it spin--structure} on a $SO(p,q)$--bundle is a
$\Spin(p,q)$--structure with respect to the covering of Lie groups
$\rho \colon \Spin(p,q) \rightarrow SO(p,q)$
(see Milnor in [8]).
For more details in Clifford--algebras and spinor groups see also [7,14,15,16].
When a spin structure is given $\eta \colon \Spin(M,g) \rightarrow SO(M,g)$,
the principal bundle $\Spin(M,g)$ is sometimes called the
{\it bundle of spinor frames}.

\noindent Let us also recall that, in this case, the local isomorphism
between $\Spin(M,g)$ and $SO(M,g)$ allows the Levi--Civita connection
to be lifted to a connection on the principal bundle $\Spin(M,g)$,
equivariant by $\Spin(p,q)$, $p+q=n$.

\noindent Let $\omega \colon T[SO(M,g)] \rightarrow \so$
be the connection one--form determined by the Levi--Civita connection
of $(M,g)$. The corresponding connection $\sigma$ on $\Spin(M,g)$

$$
\sigma \colon T[\Spin(M,g)] \rightarrow
\spin
$$

\noindent is defined by the composition:

$$
\sigma = (\rho{'})^{-1} \circ (\eta^{\ast}\omega)
\eqno (2.6)
$$

\noindent where
$\rho{'}= T_{e}\rho \colon
\spin \rightarrow \so$ is the
Lie algebras isomorphism induced by $\rho$.

{\it Spinors} are defined via a linear
representation of the spin group $\Spin(p,q)$ on a complex vector space $S$,
with $dim_{{\bf C}}S = 2^{d}$.

\noindent In fact, let
$\chi \colon \Spin(p,q) \times S \rightarrow S$ be
a {\it spinor} representation:

$$
\chi \colon \Spin(p,q) \times S\rightarrow S:
(h,s)\mapsto \chi(h,s)\equiv h \cdot s \quad .
\eqno (2.7)
$$

\noindent We recall that any representation of $SO(p,q)$ yields
a representation of its covering group, just by composition with $\rho$.
We shall call a representation of this kind
a {\it tensor representation} of $\Spin(p,q)$.
A {\it spinor representation} of $\Spin(p,q)$
(sometimes called a double representation of $SO(p,q)$) is by definition
a linear representation of $\Spin(p,q)$ which cannot be obtained from
a representation of $SO(p,q)$.

Finally, consider the vector bundle
$S(M) \equiv \Spin(M,g) \times \Co^{d}/ \Spin(p,q)$ over $M$ associated to
the principal fiber bundle $\Spin(M,g)$, where the action of $\Spin(p,q)$ on
$\Co^{d}$ is the fundamental one defined by inclusion of $\Spin(p,q)$ in
the complexified Clifford algebra $\Co l(p,q)$.
Then a spinor field $\psi$ is a section
$x \mapsto [\tilde e_{x},\Psi(\tilde e_{x})]$ of $S(M)$, where
$\Psi \colon \Spin(M,g) \rightarrow \Co^{d}$,
$\Psi(\tilde e_{x}\cdot h) = h^{-1}\cdot \Psi(\tilde e_{x})$, is an equivariant
function, i.e. an assignment of components $\Psi(\tilde e_{x})$ in $\Co^{d}$ to
each ``spinor frame" $\tilde e_{x}$.

\vskip5pt\noindent
{\bf 3 Lie derivative of Spinor Fields}

Given a spin structure
$\Spin(M,g) \rightarrow M$,
a {\it generalized spinorial transformation}
$\Phi$ of $\Spin(M,g)$
is an automorphism
$\Phi \colon \Spin(M,g) \rightarrow \Spin(M,g)$.
As discussed above $\Phi$ is a diffeomorphism of $\Spin(M,g)$ such that
$\Phi (\tilde e \cdot h) = \Phi (\tilde e) \cdot h$,
for every $\tilde e \in \Spin(M,g), \; h \in \Spin(p,q)$.
Each $\Phi$ induces a unique
diffeomorphism $\varphi \colon M \rightarrow M$
such that $\eta_{_{M}} \circ \Phi = \varphi \circ \eta_{_{M}}$,
with $\eta_{_{M}} = \pi \circ \eta$.
Moreover, in this case, $\Phi$ also induces a unique
automorphism (i.e. equivariant diffeomorphism)
$\phi \colon SO(M,g) \rightarrow SO(M,g)$
such that $\eta \circ \Phi = \phi \circ \eta$.

\noindent Equivalently we have the following diagram:
\vskip 5pt

$$
\matrix{
\Spin(M,g) & \mapright{\Phi} &\Spin(M,g) \cr
\mapdown\eta &  & \mapdown\eta \cr
SO(M,g) & \mapright{\phi} & SO(M,g) \cr
\mapdown\pi &  & \mapdown\pi \cr
M & \mapright{\varphi} & M
}
\eqno (3.1)
$$
\vskip 5pt

Furthermore, let $S(M)$ be the fiber bundle of spinors associated to the
principal bundle $\Spin(M,g)$. For every element $\Phi \in \Aut(\Spin(M,g))$
we obtain an automorphism
$\Phi_{S(M)}$ of $S(M)$, $\Phi_{S(M)} \colon S(M) \rightarrow S(M)$,
given by
$\Phi_{S(M)}(z) = [\Phi(\tilde e), s]$, $\> z=[\tilde e, s] \in S(M)$,
which is well defined because it does not depend on the representative chosen.
In fact, this is so since $\Phi$ is equivariant under the right
action of the structure group $\Spin(p,q)$.
A vector field $\tilde \Xi \in {\frak X}(\Spin(M,g))$
generating a one--parameter group
of automorphisms of $\Spin(M,g)$ defines the generalized Lie derivative
of any section
$\psi \colon M \rightarrow S(M)$, that is of any spinor field, as follows:

$$
\tilde{\cal L}_{\tilde \Xi} \psi = T\psi \circ \xi - \tilde \Xi_{S(M)}
\circ \psi
\quad ,
\eqno (3.2)
$$

\noindent where $\xi \in {\frak X}(M)$ is the only vector field such that
$T\pi \circ \Xi = \xi \circ \pi$, $\Xi$ is the only vector field such that
$T\eta \circ {\tilde \Xi} = \Xi \circ \eta$,
and for $\tilde \Xi_{S(M)}$
we have:
$\tilde \Xi_{S(M)}(z) = {d \over dt}[(\Phi_{t})_{S(M)}(z)]|_{t = 0}$, for
every $z \in S(M)$.

\noindent Once again we have the following diagram:
\vskip 5pt

$$
\matrix{
\Spin(M,g) & \mapright{\tilde \Xi} &T[\Spin(M,g)] \cr
\mapdown\eta  &  & \mapdown{T\eta} \cr
SO(M,g) & \mapright{\Xi} & T[SO(M,g)] \cr
\mapdown\pi &  & \mapdown{T\pi} \cr
M & \mapright{\xi} & TM
}
\eqno (3.3)
$$
\vskip 5pt
\noindent Hence {\it the generalized Lie derivative of} $\psi$
{\it with respect to} $\tilde \Xi$,
$\tilde {\cal L}_{\tilde \Xi} \psi \colon M \rightarrow V[S(M)]$,
takes values in the vertical sub-bundle $V[S(M)] \subset T[S(M)]$ and
being $S(M)$ a vector bundle there is a canonical
identification $V[S(M)] = S(M) \oplus S(M)$. In this case,
$(3.2)$ is of the form
$\tilde {\cal L}_{\tilde \Xi} \psi = (\psi, {\cal L}_{\tilde \Xi} \psi)$.

\noindent The first component being the original section $\psi$,
the second component
${\cal L}_{\tilde \Xi} \psi$ is also a section of $S(M)$,
and it is called the
{\it Lie derivative of} $\psi$ {\it with respect to} $\tilde \Xi$.
Sometimes, the second component  ${\cal L}_{\tilde \Xi} \psi$ is called
the {\it restricted Lie derivative} [17]. Using the fact that the second
component of $\tilde {\cal L}_{\tilde \Xi} \psi$ is the derivative of
$(\Phi_{-t})_{S(M)} \circ \psi \circ \varphi_{t}$ for $t = 0$
in the classical
sense, one can re-express the restricted Lie derivative in the form

$$
({\cal L}_{\tilde \Xi}\psi)(x) =
\lim_{t \to 0} {1 \over t}
{( (\Phi_{-t})_{S(M)} \circ \psi \circ \varphi_{t}(x) - \psi(x) )}
\quad .
\eqno (3.4)
$$

\noindent Further details and deeper discussions on the general theory
of Lie derivatives may be found in [17,18,19].

\vskip5pt
We are now seeking a coordinate expression for the restricted Lie derivative
given by the equation $(3.4)$.
Now, if $(x^{\lambda}, \psi^{i})$ denote local fibered coordinates
for $S(M)$ then $(3.4)$ reads as:

$$
{\cal L}_{\tilde \Xi} \psi = [ \xi^{a}{\bf e}_{a}\psi^{i} -
\tilde \Xi^{i}_{j}\psi^{j} ] {\bf f}_{i}
\quad ,
\eqno (3.5)
$$

\noindent where $\xi = \xi^{a}(x){\bf e}_{a}$,\
%=\xi^{\alpha}(x) e_{\alpha}^{\mu}(x)
%\partial_{\mu}$
$e = ({\bf e}_{a}) = (e_{a}^{\mu}(x)\partial_{\mu})$
is a local section
of $SO(M,g)$ induced by a local section of $\Spin(M,g)$
$\tilde e \colon U \rightarrow \Spin(M,g)$ such that
$\eta \circ \tilde e = e$, ${\bf e}_{a}(\psi^{i})$
is the Pfaff derivative of
$\psi^{i}$, that is
${\bf e}_{a}\psi^{i} = e_{a}^{\mu}(x)\partial_{\mu}\psi^{i}$ and
${\bf f}_{i}$ is a basis of $[S(M)]_{x}$ at each $x \in U$.

Recalling that $\tilde \Xi$ projects on $\Xi$ and
taking into account the Lie algebras isomorphism
$\rho{'} = T_{e}\rho \colon
\spin \rightarrow \so$,
we find the following relations:

$$\tilde \Xi^{i}_{j}= -{1\over 4}
\Xi^{a}\ida_{b}(\gamma_{a}\gamma^{b})^{i}_{j}= -{1\over 4}
\Xi_{ab}(\gamma^{a}\gamma^{b})^{i}_{j}
\eqno (3.6a)
$$
$$\tilde\Xi^{a}=\Xi^{a}=\xi^{a}
\eqno (3.6b)
$$
\noindent where indices in $(3.6a)$ are lowered and raised with respect to
the (pseudo) Riemannian metric $g$,
i.e. $\eta_{ab}=g({\bf e}_{a},{\bf e}_{b})$.
Recall that for $\Xi$, being a $SO(p,q)$-invariant vector field
of $SO(M,g)$, in the orthonormal basis $(e_{a}^{\mu}(x)\partial_{\mu})$,
we have $\Xi_{ab} = -\Xi_{ba}$. We also note that the sign
in $(3.6a)$ depends on the choice of sign for the Clifford algebra, and
the choice between $L^{a}\ida_{b}\gamma_{a}$
and $L_{b}\ida^{a}\gamma_{a}$, where
$L^{a}\ida_{b}$ and $L_{b}\ida^{a}$ are any matrices.

\noindent Finally, for the spin connection coefficients we have:

$$\sigma^{i}_{jc}= -{1\over 4}
\omega^{a}\ida_{bc}(\gamma_{a}\gamma^{b})^{i}_{j}=
-{1\over 4}
\omega_{abc}(\gamma^{a}\gamma^{b})^{i}_{j} \quad .
\eqno (3.7)
$$
So we can re--write the (3.5) as follows:

$$
{\cal L}_{\tilde \Xi} \psi = \{ \xi^{a}\nabla_{a}\psi^{i} -
[-{1\over 4}(\omega_{[ab]c}\xi^{c}
+ \Xi_{ab})
(\gamma^{a}\gamma^{b})^{i}_{j}\psi^{j} ]\,\}
{\bf f}_{i}
\quad ,
\eqno (3.8)
$$

\noindent where in the orthonormal frame $(e_{a}^{\mu}(x)\partial_{\mu})$
we have
$\omega_{abc} = - \omega_{bac}$ and
square brackets on indices mean complete
antisymmetrization, although
in this case
$\omega_{[ab]c}=\omega_{abc}$.
In fact, the coefficients $\omega_{abc}$
are the Ricci rotation coefficients defined by:

$$
\omega_{abc}= \eta_{ad}
\omega^{d}\ida_{bc} \quad .
\eqno (3.9)
$$

\noindent Moreover, for $\nabla_{a}\psi^{i}$ we have the local expression:

$$
\nabla_{a}\psi^{i} =
{\bf e}_{a}\psi^{i} + \sigma^{i}_{ja}\psi^{j} \quad .
\eqno (3.10)
$$

\noindent In this notation we have, in the same
orthonormal basis $(e_{a}^{\mu}(x)\partial_{\mu})$, the following:
$\nabla_{{\bf e}_{a}}{\bf e}_{b}
= \omega^{c}\ida_{ba}{\bf e}_{c}$ for the connection
coefficients and
${\bf\omega}^{a}\ida_{b}=
\omega^{a}\ida_{bc}(x)\theta^{c}$
for the connection one-forms, where $\theta^{c}$ is
the dual basis of ${\bf e}_{a}$. Hence, in our notation we get for
the covariant derivative $\nabla_{a}\xi_{b}$ the
following local expression:

$$
\nabla_{a}\xi_{b} =
{\bf e}_{a}\xi_{b} - \omega^{c}\ida_{ba}\xi_{c}=
{\bf e}_{a}\xi_{b} - \omega_{cba}\xi^{c}
\quad ,
\eqno (3.11)
$$

\noindent where $\xi_{c} = \eta_{cb}\xi^{b}$.

Given a space (and time) oriented (pseudo) Riemannian
manifold $(M,g)$ and a vector field $\xi$
it is possible to give its lift to $SO(M,g)$,
denoted by $\xi_{K}$. The vector field $\xi_{K}$ has the following
coordinated expression:

$$
\xi_{K} = (\xi_{K})^{a}{\bf e}_{a} +
(\xi_{K})_{ab}{\bf A}^{ab} \quad ,
\eqno(3.12)
$$

\noindent where the coefficients $(\xi_{K})^{a}$ and
$(\xi_{K})_{ab}$ are given by:

$$
(\xi_{K})^{a} = \xi^{a} \qquad
(\xi_{K})_{ab}= - \{ \nabla_{[a}\xi_{b]} +
\omega_{[ab]c}\xi^{c} \}
\quad ,
\eqno(3.13)
$$

\noindent and where $
\nabla_{[a}\xi_{b]} =
{\bf e}_{[a}\xi_{b]} -
\omega_{c}\ida_{[ba]}\xi^{c}$. The vector fields
${\bf A}^{ab}$ are local right $SO(p,q)$--invariant
vector fields on $SO(M,g)$ and in a suitable chart
$(x^{\mu},u^{b}_{a})$
are defined as follows:
$$
{\bf A}^{ab}={1\over 2}(\eta^{ac}\delta^{b}_{d}
- \eta^{bc}\delta^{a}_{d})
u^{d}_{e}{\partial\over \partial u^{c}_{e}}
\quad .
\eqno(3.14)
$$

%$e = ({\bf e}_{\alpha}) = (e_{\alpha}^{\mu}(x)\partial_{\mu})$
\noindent In order to define local coordinates $(x^{\mu},u^{b}_{a})$
which appear in $(3.14)$
we remind that, as above, the local section
$e \colon U \rightarrow SO(M,g)$ is induced by a local section of $\Spin(M,g)$
$\tilde e \colon U \rightarrow \Spin(M,g)$, so that we have
$\eta \circ \tilde e = e$. Moreover, if $(x^{\mu})$ are local coordinates
of a chart of the base manifold $M$ with the same domain $U$, we can define
local coordinates $u^{\mu}_{a}$
by $(u_{a}^{\mu}(u)\partial_{\mu})$
where $u = ({\bf u}_{a})$ is any frame
in the open $\pi ^{-1}(U)$ of $L(M)$ such that
$g({\bf u}_{a},{\bf u}_{b})=\eta_{ab}$.
Finally, with the local section $e \colon U \rightarrow SO(M,g)$
we define local coordinates $u^{b}_{a}$
by ${\bf u}_{a}=u^{b}_{a}{\bf e}_{b}$.

\vskip5pt
\noindent Accordingly, $\xi_{K}$ transforms as a $SO(p,q)$--invariant
vector field on $SO(M,g)$ and projects over $\xi$.
An alternative geometric definition of $\xi_{K}$ is the following one.
Given any vector field $\xi$ of the base manifold $M$,it admits
a unique natural lift $\hat\xi$ to the linear frame bundle $L(M)$.
Then, $\xi_{K}$ is by definition the skew of $\hat\xi$
with respect to the (pseudo) Riemannian metric $g$.
\vskip5pt
\noindent Furthermore, when a spin structure is given
$\eta \colon \Spin(M,g) \rightarrow SO(M,g)$, any vector field
$\xi$ lifts to a vector field on the spin bundle
$\Spin(M,g)$. In fact, its lift $\tilde\xi_{K}$ is defined by:

$$
\tilde\xi_{K} = (\tilde\xi_{K})^{a}{\bf e}_{a} +
(\tilde\xi_{K})^{i}_{j}{\bf E}^{j}_{i} \quad ,
\eqno(3.15)
$$

\noindent and the coefficients $(\tilde\xi_{K})^{a}\>$,
$(\tilde\xi_{K})^{i}_{j}\>$ are given by:

$$
(\tilde\xi_{K})^{a} = \xi^{a} \qquad
(\tilde\xi_{K})^{i}_{j}= -{1\over 4}(\xi_{K})_{ab}
(\gamma^{a}\gamma^{b})^{i}_{j} \quad ,
\eqno(3.16)
$$

\noindent so that $\tilde\xi_{K}$ projects over $\xi_{K}$ and
hence over $\xi$.
Accordingly, the following diagram commutes:
\vskip 5pt
$$
\matrix{
\Spin(M,g) & \mapright{\tilde \xi_{K}} &T[\Spin(M,g)] \cr
\mapdown\eta&  & \mapdown{T\eta} \cr
SO(M,g) & \mapright{\xi_{K}} & T[SO(M,g)] \cr
\mapdown\pi &  & \mapdown{T\pi} \cr
M & \mapright{\xi} & TM
}
\eqno (3.17)
$$
\vskip 5pt
The vector field $\tilde\xi_{K}$ is well defined because the one--parameter
group of automorphisms $\lbrace\phi_{t}\rbrace$ of $\xi_{K}$
can be lifted (see below) to an one--parameter group of automorphisms
$\{\tilde\phi_{t}\}$ so that
$\tilde\xi_{K}(\tilde e) = {d \over dt}[\tilde\phi_{t}(\tilde e)]|_{t =
0}$, for
every $\tilde e \in \Spin(M,g)$.

The vector fields $\tilde\xi_{K}$, $\xi_{K}$,
invariant under the actions of $\Spin(p,q)$ and $SO(p,q)$,
are called the
{\it Kosmann vector fields} of the vector field $\xi$. Let us also remark
that the Kosmann lifting $\xi \longmapsto \xi_{K}$ is not
a Lie--algebra homomorphism although the following remarkable relation holds:
$$
[\xi,\zeta]_{K} = [\xi_{K},\zeta_{K}] -
{1\over 2}{\cal L}_{\xi}(g_{\lambda\mu})
g^{\mu\nu}{\cal L}_{\zeta}(g_{\nu\sigma})
{\bf A}^{\lambda\sigma}
\eqno (3.18a)
$$

\noindent where

$$
{\bf A}^{\lambda\sigma}=e^{\lambda}_{a}e^{\sigma}_{b}{\bf A}^{ab}
\quad .
\eqno (3.18b)
$$

\vskip5pt
\noindent We are now in a good position to state the following result.

\vskip5pt
\noindent{\bf Proposition}. {\it The Lie derivative
${\cal L}_{\xi}\psi$ of a
spinor field $\psi \colon M \rightarrow S(M)$ with respect to a vector
field $\xi \in {\frak X}(M)$, as defined by Kosmann in [9],
is the second component of the generalized Lie derivative of $\psi$
with respect to the Kosmann lift $\tilde\xi_{K}$ of $\xi$.
That is if
$\tilde{\cal L}_{\tilde \xi_{K}}\psi$ is
the generalized Lie derivative of $\psi$ and
${\cal L}_{\tilde \xi_{K}}\psi$ its second component then
we get
${\cal L}_{\xi}\psi = {\cal L}_{\tilde \xi_{K}}\psi$.}

\vskip5pt
\noindent{\bf Proof:}  It is an immediate consequence of
the above considerations and from the fact that the local expression
of the Lie derivative given by Kosmann is
$$
{\cal L}_{\xi}\psi = [ \xi^{a}\nabla_{a}\psi^{i}
-{1\over 4}\nabla_{[a}\xi_{b]}
(\gamma^{a}\gamma^{b})^{i}_{j}\psi^{j} ]
{\bf f}_{i}
\quad .
\eqno (3.19)
$$
\rightline{(Q.E.D)}
\vskip5pt
\noindent We remark that the above proposition gives us a geometric meaning
to the Kosmann's formula $(3.19)$ which is taken by Kosmann as a definition
of the Lie derivative of $\psi$ with respect to $\xi$.
We shall use the following
notation: $\tilde{\cal L}_{\xi}\psi := \tilde{\cal L}_{\tilde \xi_{K}}\psi$.
We finally remark that there is an easier interpretation of
the above facts in the case in which
$\xi$ is an infinitesimal isometry, i.e. a Killing vector field on $M$.

In fact, given a spin structure $\eta \colon \Spin(M,g) \rightarrow SO(M,g)$,
let us consider a Killing vector field $\xi$, its flow denoted by
$\lbrace\varphi_{t}\rbrace$
and the principal bundle map
$\phi_{t} \colon SO(M,g) \rightarrow SO(M,g)$ which is defined
by restriction of $L_{\varphi_{t}} \colon L(M) \rightarrow L(M)$ to $SO(M,g)$.
The automorphisms $\phi_{t}$ are well defined since the
diffeomorphisms $\varphi_{t}$ are isometries.
Moreover being $\eta \colon \Spin(M,g) \rightarrow SO(M,g)$
a covering space it is
possible to lift $\phi_{t}$ to a bundle map
$\tilde\phi_{t} \colon \Spin(M,g) \rightarrow \Spin(M,g)$ in the following way.
For any spinor frame $\tilde e \in \Spin(M,g)$ over $\eta (\tilde e)=e$,
from the theory of covering spaces it follows that for the curve
$\gamma_{e} \colon \Re \rightarrow SO(M,g)$, based at $e$, that is
$\gamma_{e}(0)=e$, and defined by $\gamma_{e}(t):=\phi_{t}(e)$ there
exists a unique curve
$\tilde \gamma_{\tilde e} \colon \Re \rightarrow \Spin(M,g)$,
based at $\tilde e$,
such that $\eta \circ \tilde \gamma_{\tilde e}=\gamma_{e}$. It is possible
to define a principal bundle map
$\tilde\phi_{t} \colon \Spin(M,g) \rightarrow \Spin(M,g)$, covering $\phi_{t}$,
by letting $\tilde\phi_{t}(\tilde e):=\tilde \gamma_{\tilde e}(t)$.
The one--parameter group
of automorphisms of $\Spin(M,g)$
$\{\tilde\phi_{t}\}$ defines a vector field
$\tilde \Xi (\tilde e) = {d \over dt}[\tilde\phi_{t}(\tilde e)]|_{t = 0}$, for
all $\tilde e \in \Spin(M,g)$.

\noindent It is then easy to verify that
the vector field $\tilde \Xi$ coincides with the
Kosmann vector field $\tilde\xi_{K}$ of the vector field $\xi$.
In fact, in this case, the lift $\hat\xi$ is already skewsymmetric.
Hence the Lie derivative of $\psi$ with respect to $\xi$ is still given by
the Kosmann's formula $(3.19)$ (cf. [23,24])
or, better, by the following
coordinate--free expression

$$
({\cal L}_{\xi}\psi)(x) =
\lim_{t \to 0} {1 \over t}
{( (\tilde\phi_{-t})_{S(M)} \circ \psi \circ \varphi_{t}(x) - \psi(x) )}
\quad .
\eqno (3.20)
$$

Moreover, even for an arbitrary vector field $\xi$ the formula $(3.20)$
is still valid but where now $\lbrace\phi_{t}\rbrace$ is the flow of
the vector field $\xi_{K}$, and $\lbrace\tilde\phi_{t}\rbrace$
is the flow of $\tilde\xi_{K}$.
Now, $\xi$ is not an infinitesimal isometry but $\xi_{K}$ generates
a one--parameter group of principal automorphisms of $SO(M,g)$ which
lift to $\Spin(M,g)$ so that even in this case $(3.20)$ gives us a further
interpretation of Kosmann's formula, i.e. a direct classical definition
of the Lie derivative for spinor fields.

\vskip5pt\noindent
{\bf 4 Acknowledgments}

We are deeply grateful to A. Borowiec and G. Magnano for
their useful remarks.
One of us (M.G)
is greatly indebted to G. Chaix from Paris for his editorial assistance
during the preparation of this work.
This work is sponsored by GNFM, MURST (40\% Project ``Metodi Geometrici
e Probabilistici in Fisica Matematica'').

\vskip5pt\noindent
{\bf References}

\noindent[1]
		 Dirac, P.A.M., (1928),
			{\it The Quantum Theory of the Electron}, Proc. Roy. Soc. of Lond. A.
			{\bf 117} 610--624.

\noindent[2]
		 Dirac, P.A.M., (1958),
			{\it The Principles of Quantum Mechanics}, Clarendon Press,
			Oxford (1930) fourth edition.

\noindent[3]
		 Pauli, W., (1927),
			{\it Zur Quantenmechanik des magnetischen Elektrons},
				Z. Phys. {\bf 43} 601.

\noindent[4]
		 Cartan, E., (1913),
			{\it Les groupes projectifs qui ne laissent invariante aucune
				multiplicit plane},
				Bull. Soc. Math. France {\bf 41} 53--96.

\noindent[5]
		 Chevalley, C., (1954),
		 {\it The Algebraic Theory of Spinors}, Columbia Univ. Press.

\noindent[6]
		 Dieudonn\'e, J., (1963),
			{\it La g\'eometri\'e des groupes classiques},
			deuxi\`eme \'edition, Springer--Verlag, Berlin.

\noindent[7]
		 Atiyah, M. F., Bott, R., Shapiro, A., (1964),
		 {\it Clifford modules}, Topology {\bf 3}, Suppl. 1, 3--38.

\noindent[8]
		 Milnor, J., (1963), Enseignement Math. (2) {\bf 9}, 198--203.

\noindent[9]
		 Kosmann, Y., (1972), Ann. di Matematica Pura et Appl. {\bf 91} 317--395.

\noindent[10]
		 Kosmann, Y., (1966), Comptes Rendus Acad. Sc. Paris,
			s\'erie A, {\bf 262} 289--292.

\noindent[11]
		 Kosmann, Y., (1966), Comptes Rendus Acad. Sc. Paris,
			s\'erie A, {\bf 262} 394--397.

\noindent[12]
		 Kosmann, Y., (1967), Comptes Rendus Acad. Sc. Paris,
			s\'erie A, {\bf 264} 355--358.

\noindent[13]
		 Greub, W., Petry, H.R., (1978), {\it On the lifting of structure groups},
			in Lecture Notes in Mathematics {\bf 676}, 217, Springer--Verlag, NY.

\noindent[14]
		 Blaine Lawson, H., Jr., Michelsohn, M.-L., (1989), {\it Spin Geometry},
			Princeton University Press, New Jersey.

\noindent[15]
		 Dabrowski, L., (1988), {\it Group Actions on Spinors}, Lectures
			at the University of Naples, Bibliopolis, Napoli, Italy.

\noindent[16]
		 Slov\'ak, J., (1992),
				{\it Invariant Operators on Conformal Manifolds},
				Lecture series at the University of Vienna, Fall term 1991/1992
				(to be published).

\noindent[17]
		 	Kol{\'a}{\v r}, I., Michor, P.W., Slov\'ak, J., (1993),
				{\it Natural Operations in Differential Geometry},
				Springer--Verlag, NY.

\noindent[18]
		 	Jany{\v s}ka, J., Kol{\'a}{\v r}, I., (1982), in
				{\it Proceedings of the
	   Conference on Differential Geometry and its applications,
				September 1980}, (ed. Kowalski O.), Published by Univerzita Karlova Praha,
				pp. 111--116.

\noindent[19]
				Salvioli, S.A., (1972), {\it On the theory of geometric objects},
				J. Differential Geometry, {\bf 7} 257--278.

\noindent[20]
				Garc\'\i a, P.L., (1972), {\it Connections and 1--jet Fiber Bundles},
				Rend. Sem. Mat. Univ. Padova, {\bf 47} 227--242.

\noindent[21]
		 	Eck, D.J., (1981),
				{\it Gauge--natural bundles and generalized gauge theories},
					Mem. Amer. Math. Soc. {\bf 33} {\bf No 247}.

\noindent[22]
		 		Haefliger, A., (1956), {\it Sur l'extension du groupe structural d'un
						espace fibr\'e},
					Comptes Rendus Acad. Sc. Paris, {\bf 243} 558--560.

\noindent[23]
		 		Lichnerowicz, A., (1963),  {\it Spineurs harmoniques},
					Comptes Rendus Acad. Sc. Paris, groupe 1, {\bf 257} 7--9.

\noindent[24]
				Dabrowski, L., (1993), in {\it Spinors, twistors, Clifford algebras,
				and quantum deformations}, (edited by Oziewicz, O., Jancewicz, B.,
				and Borowiec, A.), Kluwer Academic Publisher, The Netherlands, pp. 61--65.

\end